# A Unity of Science, Especially Among Physicists, Is Urgently Needed to End Medicine's Lethal Misdirection

John T. A. Ely, Ph.D.
Radiation Studies, Box 351310
University of Washington
Seattle WA 98195

**ABSTRACT**
We have all read that: (1) organized medicine was laughing at the germ theory and refused to wash its hands in the late 1800's while women died of childbed fever and other patients of wound sepsis (scientists in general, including physicists, were among those who died); and (2) during the 1900's, and still today, organized medicine is laughing at the essential nutrient theory. We now hear the medical quality assurance boards enforce MANY MISTAKEN VIEWS and DO NOT recognize that: "xenobiotics can't be substituted for essential nutrients; ascorbic acid is NOT a vitamin and hence, is needed in multi-gram (not minute amounts) for optimum health; there is great harm in a diet deriving half its calories from refined carbohydrates as is common in the US today; dietary correction is possible at a stage when surgery is contemplated for a cardiac patient; mercury dental fillings and even the high-copper variety most widely used are extremely unsafe;" etc. The purpose of this manuscript is to elicit coordinated collective effort of physicists, other scientists and the Union of Concerned Scientists to educate the public and end this tyranny of organized medicine whose methods are reported to cause much M&M (over a million unnecessary US deaths at a cost of many billion dollars annually). The biochemistry of these errors would be trivial to understand for physicists and other scientists (who are all still going innocently to slaughter among those dying needlessly, even today).

**List of Abbreviations**
AA = ascorbic acid, ascorbate
AGC = aggressive glycemic control
CDC = Centers for Disease Control
cmi = cell mediated immunity
cvd = cardiovascular disease
GAA = glucose ascorbate antagonism
GHb = glycated hemoglobin
g/d = grams per day
HMS = hexose monophosphate shunt
iv = intravenous
LANL = Los Alamos National Laboratory
M&M = morbidity and mortality
mg% = mg/dL = mg/deciliter = mg percent
rCHO = refined carbohydrate
TOC = treatment of choice
UCS = Union of Concerned Scientists
wbc = white blood cells
WHO = World Health Organization

   See newly added Chapter 2 on mercury amalgams and Extreme Psychoses (Serial Killers with No-Remorse) beginning on page 18. Note that in what follows, some repetition is necessary so that individual sections can be read separately.

# BACKGROUND

The rationale for publishing this paper on LANL arXiv relates to the roles, responsibilities and duties of physicists to protect the society that supports them. When something new and important happens in physics every physicist wants to hear about it. When something new and important happens in medicine, almost no one has time to hear about it. We don't want to villianize physicians; treatments of choice that substitute xenobiotics for essential nutrients are reportedly so ineffective that vast numbers of patients are dying unnecessarily and physicians cannot take time to indulge their natural interest in medicine. If instead, essential nutrient therapies were used, ending the carnage, so few people would be ill that most physicians would be very lightly burdened in their practice, which would have become essentially preventive medicine. Thus, the purpose of this message will have been accomplished.

**Poor Health Status in Nations Using US Medical System**
World Health Report 2000 (WHO 2000), issued in June by the World Health Organization, ranks 191 United Nations (UN) member states on the quality, cost, etc., of their health care systems. Although the U.S. spends far more than any other nation it is ranked only 37th! Countries with medical systems most similar to the U.S. also scored very poorly: Australia 32nd, Canada 30th and New Zealand 41st. The probability of these four nations grouping so closely (ie, in a span of only 12 of 191 rankings) by chance is extremely low, $p<0.001$ [ie, $(12/191)^{**}3$]. Moreover, in spite of the U.S. medical expenditure being far more than any other nation, it is ranked 72nd in personal health and NZ, spending more than circa 90% of all other nations, is ranked 80th. It is improbable that these two countries would rank so closely by chance (ie, in a span of only 8 of 191 rankings, $p<0.05$). Our analysis herein shows that the low rankings earned by US style medical systems is due almost entirely to impact of errors in modern medicine (i.e., xenobiotics have been substituted for essential nutrients which are rejected as unprofitable). It is shown below that the vast majority of NZ and US deaths, suffering and incapacitation, and cost also result from the same errors, of which ignorance regarding hyperglycemia and the "mistaken view " that ascorbic acid (AA) is a vitamin are among the most culpable.

In the US, it has been virtually canonical that the practice of medicine is and should be a prestigious profession. In recent decades of health care quality decay cited by highly respected authorities (Watson 1973; Relman 1992; Angell 2000), physicians of our acquaintance, as individuals, have been highly trained and motivated (no one works harder), and would have welcomed new more effective modalities, even if unprofitable. Yet, all of medicine has been losing respect, including individual physicians through no fault of their own. Relevant are public expectations of physicians that are not humanly possible to fulfill, e.g., they are expected to know what is known (i.e., published in the medical science literature). This knowledge has become essentially inaccessible because of its size and growth rate. A large university medical center library receives over 40 million new pages per decade from the journals to which it subscribes. Much of this material is both excellent and relevant to clinical medicine. However, it isn't possible even to turn this many pages in a decade. No clinician (or even researcher) could possibly find the time to locate, by any system, and read 1% (40,000 pages/year) of the flood. There may be no area of human endeavor where ignorance (of what is known) is growing as rapidly as in clinical medicine.

Ignorance, however, is no excuse for breaking the law. In medicine, the law is derived from the literature. As demonstrated here: (1) the "complete" literature (i.e., ~1935 to date) contains all that is needed to cure most of the major diseases with already proven and easily tested modalities; and (2) scientific study of these modalities actually defines the laws of nature with respect to medicine. However, the literature also contains vast reams of publications apparently written by authors who have ignored the "complete" literature, and are ignorant of the laws of nature. These authors advocate empirical and entirely unscientific, often unsafe and inefficacious outmoded treatments called treatments of choice (TOC's) which are vigorously enforced by medical disciplinary boards. This has controlled medicine and prevented its modernization by modalities that were unpatentable and therefore rejected by certain interests as unprofitable (ie, in effect, analogous to opposing hand washing ~1850AD). Thus, TOC's are causally associated with the extremely high US M&M and health care costs, and very low WHO rankings. There are 200,000 deaths/yr ($18 billion/yr) due to infections; 500,000 due to cancer; and 700,000 due to cardiovascular disease. The total annual bill for the "care" producing this toll is reported to be > $1 trillion (> 1/6 of the US national debt) (Feltman 1991; CDC 2000). In addition to this bankrupting cost, it appears that as many people die in the U.S. each year due to failure of TOC's as were executed annually in Hitler's death camps. It appalls to hear the expression "Death Camp USA".

In Europe between 1840 and 1870, studies conducted by Ignaz Semmelweis, John Snow, Louis Pasteur, and Joseph Lister, contributed to the growing science of bacteriology (Lederberg 2000). The germ theory was slow to be transmitted to the United States, however. We have all read that U.S. physicians laughed at the germ theory and refused to wash their hands in the late 1800's while women died of childbed fever and other patients of wound sepsis. It was not until the late 1880s that antiseptic surgery was started in the US by American surgeon Dr. William Mayo in his clinics in Rochester, Minnesota. There is evidence that some senior physicians in U.S. hospitals still had not accepted the germ theory circa 1900 (Price 1917). History insists medicine has shown the same characteristic behavior, in the 1700's denying that blood circulated in the body, then denying the germ theory, and recently, in the 20th century denying basic biochemical knowledge regarding the science of essential nutrients including that of AA. Thus, medicine has been accused of being both the sole judge of its own quality and historically calloused in its slowness to modernize by elimination of long held erroneous views. In each era, these views are reported to vigorously oppose basic facts readily proven from the literature to be of great importance and subsequently adopted. Even now, these views are reported to cause immeasurable unnecessary suffering and mortality, including a million early deaths in treatment annually in the U.S. (Feltman 1991; CDC 2000).

The similarity between "laughing at the germ theory" and rejecting modalities as unprofitable (in spite of higher safety and efficacy) is extremely important. Pauling called such behavior "condemnation without investigation". Medicine has an irrevocable sworn duty to act in the best interests of the patient. Isn't it criminal either to adopt or continue use of therapies with low safety and efficacy, many side effects and high cost (especially when profit is a motive)? We discuss below a sampling of such instances (from a much larger number that exist) and ask how can the resulting M&M be justified? It is also puzzling that medicine ignores some matters with more negative impact on human health than did ignorance of the germ theory.  These discussed in later sections, include obvious gross changes in glycemic levels in the affluent societies since 1900, and certain biochemical problems of humans elucidated by study of other mammals.

Of course, it is clear that many thousands of exemplary physicians have humanistic motives and exert themselves to heroic levels, accomplishing much for their patients.  However, the complete literature argues that, if the "rejected modalities" recommended here were employed, M&M and costs would be far lower in the hands of all clinicians, including the large number that are already exemplary.  Illustrative examples are Denham Harman, MD, PhD and Abram Hoffer, MD, PhD.  As explained (inter alia): (1) Harman showed that both "normal" aging and the major life shortening mitochondrial aging could be significantly slowed if the "rejected" antioxidants (ascorbate, ubiquinone and vitamins A and E) were accepted and properly used; and (2) Hoffer's early work with essential nutrients prompted Linus Pauling to coin the term "orthmolecular" (Pauling 1969).

**More on the Decline of U.S. Health Care Quality**
The deterioration in personal health from circa 1950 to 1970 was striking and quantifiable, but not understood because of fundamental errors in the canonical views.  To establish some important temporal aspects of this decay, in addition to WHO, we cite six prominent persons (Watson 1973; Relman 1992; Angell 2000; Stevens and Glatstein 1996); Trowbridge 1986):

The first is the following statement delivered at the Mayo Clinic in Rochester, Minnesota, on 19 November 1970, by Mr. Thomas J. Watson, Jr., CEO of IBM:

"Let me start by asking a question that this great medical center brings to mind: How would you like to live in a country which, according to the figures available in the United States during the past two decades: has dropped from seventh in the world to sixteenth in the prevention of infant mortality; has dropped in female life expectancy from sixth to eighth; has dropped in male life expectancy from tenth to twenty-fourth; and which has bought itself this unenviable trend by spending more of its gross national product for medical care ($1 out of every $14) than any other country on the face of the earth?"

"You know the country I am talking about: Our own U.S.A., the home of the free, the home of the brave, and the home of a decrepit, inefficient, high-priced system of medical care. Just look for a moment at what some of the figures mean. They mean that in infant mortality we have been overtaken by France, the U.K., and Japan, that in male life expectancy we have been overtaken by France, Japan, West Germany and Italy. I know experts can disagree over our precise international standing. And I realize that medical problems in the United States, Europe and Japan are not identical. But the evidence overwhelmingly indicates that we are falling down on the job, heading in the wrong direction, and becoming as a nation a massive medical disgrace."

The second is the paragraph below from a 1992 article on health-care reform by Arnold S. Relman, MD, long-time Editor-in-Chief of the New England Journal of Medicine (NEJM):

"In short, our health-care system, formerly a social service that was the responsibility of dedicated professionals and not-for-profit facilities, has become a vast, profit-oriented industry. The revenue of this industry constitutes the country's  health-care costs.  As in any other industry, providers constantly strive to increase their profitable sales, but unlike other industries, consumers exercise little control over their consumption of products and services. It should not be surprising that such a system is afflicted not only with relentless inflation but also with neglect of the needs of the uninsured and with failure to promote the use of valuable but unprofitable health services."

The third is an editorial by Marcia Angell, MD, another former NEJM Editor-in-Chief, entitled "Is Academic Medicine for Sale?" She discusses numerous involved facets of industry influence that actually affect the fundamental views of physicians and the way they practice medicine (i.e., "As the critics of medicine so often charge, young physicians learn that for every problem, there is a pill ..."). In support of the Angell editorial's thesis, as epitomized by this quote, we show below examples of the countless tragedies dominating medicine because of the simplistic "pill" mindset that is unmindful of homeostasis, the literature and the essential nutrients.

Angell's editorial strongly supports, essentially verbatim, a warning made in 1983 by the fourth, immunologist Alan S. Levin, MD, JD (Trowbridge 1986). His medical symposium presentation analyzed problems due to industrial control of medicine. He pointed out that a sample issue of possibly the world's largest and most prestigious medical journal had 51 pages of scientific matter and 99 pages of industry ads. He said: industry is setting the standard of medical practice and has invaded the academic institutions; as much as two-thirds of the research funding in most major institutions comes directly or indirectly from industry interests; therefore, chemotherapy is dominant, and, unfortunately, immunotherapy is secondary. He added that: the market for (sorely needed) immunotherapy in the US is only $50 million while the market for (heavily advertised) antihistamines is about $2 billion; medicinals are more profitable than any other industrial product; the practice of medicine and our medical science are dictated by industry that is promoting its own financial welfare. His talk ended with the following excerpt:

". . . this is no laughing matter . . . We are being taken over by . . . industry. It is using the academic institutions to do this. If you don't think so, read any of the so-called peer-review journals, and you will realize that they are dominated by . . . industry."

It has been widely asserted that the matters most vital to the people including the expense, pain and sorrow endured and their very survival have been systematically and relentlessly subordinated to business interests via influence on the quality and cost of medical care ((Watson 1973; Relman 1992; Angell 2000; Stevens and Glatstein 1996; Trowbridge 1986; Pauling 1969, 1987). The interests refuse to recognize even the tragic failure of the U.S. medical system, let alone existence of possible errors in accepted views of biology. Finally, in an editorial, Stevens and Glatstein (1996) have emphasized that our failures to cure the major diseases will not yield to more extravagant expenditures if we do not understand the biology:

"Beware the Medical-Industrial Complex.
'We must guard against the acquisition of unwarranted influence, whether sought or unsought, by the military industrial complex.' Farewell Address of Dwight D. Eisenhower, 34th President of the US, January 1961. If Ike were with us today, he might well expand his views on power and influence to include modern American medicine. . . Our true challenge . . . is to *understand the biology* . . . We must not be . . another special interest come to drink at the well of public spending, but as advocates for the public good. If we fail . . . we will be unable to make any important long-term contribution to those who matter most - our patients."

**SECTION 1.  Glycemic Modulation in Health and Disease.**

   Cures of most major disorders have in common the need to keep blood glucose in a low range that would be normal among people of undeveloped nations who have an unrefined diet.  In his 1969 paper in Science, Linus Pauling coined the term "orthomolecular" to describe the interrelationships of molecules normally present in the body.   He was largely motivated by the work of Abram Hoffer, MD, PhD, who was able to correct a number of disorders including schizophrenia that had been considered incurable. An orthomolecular lesion common to many, possibly most, diseases exists when blood glucose is high and ascorbic acid is low.

The advent of cheap energy in affluent countries occurred in the late 1800's and enabled them to refine foods using fossil fuels. The costs of harvesting and processing food became steadily lower.  However, there was no knowledge of the effects of food refinement on its essential nutrient content in the early 1900's.  Within two decades after the introduction of low cost, shelf stable, refined foods (i.e., white flour, sucrose, etc), the first evidence of heart disease appeared .  By mid-century heart disease had become the most common cause of death in affluent societies.

An important study by Cheraskin and Ringsdorf (1974) puts all of these facts in perspective.  They cite US DA statistics that, in 1970, the average American consumed 264 pounds of empty calories.  On a dry weight basis, this was just over 50% of food intake.  Using the Cornell Medical Index Health Questionnaire (CMIHQ) and a sample of 715 well educated people, they created plots of refined carbohydrate  (rCHO) intake vs health problems.  Extrapolations of these plots show that zero health problems occur at zero rCHO intake.  Consider the implication for staggering increases in M&M and dollars that have been caused by rejection of the glycemic control modality provided by an unrefined diet.

As stated above, the literature on agriculture, food science, nutrition, and medicine has now become almost inaccessible because of its size and growth rate.  Making the wisest decision in terms of what to do with food, food processing, or essential nutrient supplementation is not trivial. Being sufficiently informed to make advancements in the struggle against the effects of fundamental errors is difficult. The general public, lawmakers, and researchers must be aware of the errors of affluence and their many influences on health.

Very few can understand the complexities and interactions of food chemistry, nutrition, and human disease. A leading exception is the originator of the free radical theory of aging, Denham Harman (1969, 1972, 1981), whose chemistry doctorate preceded his degree in internal medicine. Harman showed that both 'normal' aging and the major life-shortening mitochondrial aging could be retarded if the essential nutrient antioxidants ascorbate, ubiquinone (Coenzyme Q10), and vitamins A and E were properly used. Indeed, we argue that, if adequate levels of all the essential nutrients were present in the diet, M&M and their associated costs would be far lower.

In this era, medicine's opposition to modernization has been directed toward essential nutrients, a class of modalities that are well defined in the literature but were rejected by certain interests as unprofitable in the 20th century. Note that, in effect, rejecting the importance of essential nutrients is analogous to opposing hand washing while laughing at the germ theory a century earlier (Lederberg 2000; Price 1917). We argue the misconceptions cited above as the cause of the erroneous view on these modalities has persisted and subtly permeated all aspects of health care, including medical school textbooks. The deterioration in the quality of health from 1950 to 1970 was striking in America (despite its prosperity) and quantifiable, but not understood because of the widely accepted indifference to the theory of essential nutrients.

**Essential Nutrients, Bioenergetics and A Fundamental Error.**

Basic energy requirements of cells such as the deformation and motility needs of white blood cells (wbc) are met by ATP synthesis in the mitochondria where the essential nutrient ubiquinone is required for oxidative phosphorylation. In addition, mitosis in any tissue requires ribose to copy DNA. This is supplied by the hexose monophosphate shunt (HMS), most details of which have been well known in a number of tissues (fetal, wbc, etc) since the 1960's. In wbc, both ribose for mitosis of lymphocytes and $H_2O_2$ for phagocytes are obtained via the HMS. In 1971, the discovery by researchers at The Bowman Gray School of Medicine that HMS rate is *proportional to intracellular ascorbic acid* (AA) should have changed *all* fields of medicine (Cooper et al 1971). Because wbc were used for this work (instead of fetal or etc.), at least the immunologists should have immediately changed their thinking on AA. Certainly, most of them read "Infection and Immunity". In 1971, immunologists should have realized explicitly that intracellular AA, as measured by buffy coat (wbc) AA: (1) gives a prompt indication of immune status; and (2) is a *universal limiting factor* that determines the rate and intensity of cell mediated immune response (cmi). In humans and in AA-synthesizing mammals, lymphocytes multiply more rapidly and phagocytes ingest and kill more effectively *when their intracellular AA,* is high (as well as certain vitamins that, like AA, are also antioxidants, such as vitamins E and B6) (Grimble 1997; Penn et al 1991). In this era, it is as absurd as laughing at the germ theory to cite a Recommended Dietary Allowance (such as ~60 mg/day) for AA (called vitamin C although *not* a vitamin). The ~50mg/kg body weight synthesized daily by all 4000 normal mammals in the unstressed state is the actual requirement for AA in humans and all other non-synthesizing mammals (Pauling 1987). Possibly, the most lethal misconception of our time is the notion that AA is a vitamin. This notion is a major factor in the vast M&M cited by WHO 2000.

Thus, the assumption that the refined diet is adequate and that essential nutrients can be omitted safely, without serious or fatal consequences, is profoundly flawed. This flawed belief is especially striking for AA and for ubiquinone, and has similar but less severe consequences for other essential nutrients.

**Glycemic Modulation of the HMS.**

Three decades ago, after reading the Bowman-Gray 1971 report, we deduced and related to Linus Pauling a theoretical reason for the failure of clinical trials of AA against colds and cancer—the high blood glucose levels in developed nations. This theory is called "Glucose-Ascorbate Antagonism" (GAA). We extended this theory experimentally and obtained supporting findings in aging, birth defects, cancer, infectious diseases, and others (Ely 1981, 1996a,b; Santisteban et al 1985; Hamel et al 1986; Ely et al 1988; Ely and Krone 2002,a,b; Krone and Ely 2001, 2004) (see Section 3 below). A very high world wide correlation of cancer incidence with sugar consumption has been reported (Hems 1978; Carroll 1977; Michaud et al 2002; Nilsen and Vatten, 2001; Schoen et al 1999; Augustin et al 2003). Our theory, supported by human and animal studies, suggests that the most common factor increasing tumor tolerance in developed nations is the gross elevation of mean blood glucose (Sanisteban et al 1985; Ely 1996a). A prominent example of the diseases of affluence is the estimate that about one-fourth of those alive today in these nations will present with clinical cancer during their lifetime. The GAA predicts that such individuals can be identified approximately as the older half of the upper half by sugar consumption. The basis for this prediction can be seen in two more statistics of the affluent societies: (1) ~30 times more cancer occurs above age 55 than below 35; and (2) the 2-hour value on glucose tolerance tests rises 10% per decade of age (Ely 1996b; Ely and Krone 2002a).

We sometimes think of the HMS as paradoxical because its rate varies inversely with the concentration of its substrate, glucose. The GAA is important in the normal (low-dose) range of ~10 g/day or less AA. The theory states that the intracellular AA levels in certain cell types, such as leukocytes and fetal cells, are 'pumped up' by insulin-mediated active transport to the measured concentrations, ~50 times higher than in the surrounding plasma. This increase of intracellular AA occurs effectively if blood glucose is in the low range—50 to 90 mg/deciliter (mg/dL). This blood sugar was considered normal until the 1900s and is still seen where a primitive (unrefined) diet prevails, but it is approximately half the glycemic levels typical of affluence (Ely 1996a; Chatterjee and Bannerjee 1975).

Leukocytes require high AA levels to drive the HMS to supply adequate hydrogen peroxide and ribose for effective phagocytosis and mitosis. In even 'modest' blood glucose elevations (over ~130 mg/dL, common following most western diet meals), blood glucose molecules so outnumber AA that they competitively inhibit insulin-mediated active transport of AA into cells (Hutchinson et al 1983). Such inhibition results in low intracellular AA levels, low HMS, and cell dysfunction (namely, leukocytes don't attack tumors or pathogens, fetal cells divide too slowly, and so on); this phenomenon is the glucose 'antagonism' of AA. A principal cause of cardiovascular disease (cvd) is hyperglycemia, which reduces AA to scorbutic levels in vascular intimal cells. The GAA theory gives rise naturally to the use of controlled hypoglycemia concurrent with AA in the low-dose range (below 10 g/day). This modality is called Aggressive Glycemic Control (AGC) and appears to have much value against many disorders. In the high-AA dose range, it can out-compete glucose for insulin and elevate intracellular AA, accelerating the HMS.

Ignorance of GAA and AGC was most likely responsible for the unimpressive showings of the three large scale North American cold trials. This ignorance, certain other misperceptions, and deliberate misrepresentations, all well-cataloged and clarified by Pauling (1987), are also responsible for many erroneous views on AA, and much M&M.

**On Controlled Hypoglycemia.**

Cancer, infections and other diseases (cvd, etc) have been shown to exhibit a much lower incidence and more rapid remission with adequate AA, and with blood glucose in the 50 to 90 mg/dL range (Murata et al 1982; Riordin et al 1985; Rivers 1989). Such glycemic control can be achieved by conforming to an unrefined diet, moderate caloric restriction, and as much physical activity as ones physician approves. In insulin-coma therapy, formerly used in the treatment of psychiatric disorders, blood glucose was maintained circa 30 mg/dL, serendipidously maximizing intracellular AA. In this therapy, remissions from cancer were reported to occur in patients whose incurable malignancies were unknown to the psychiatrist at the time of treatment (Koroljow 1962). A more practical approach for AGC, avoiding the difficulties and dangers of coma, might be to maintain blood glucose at 50 to 60 mg/dL with exogenous insulin or Orinase (a non-halogenated oral hypoglycemic).

## SECTION 2. Ascorbic Acid Is Not a Vitamin

**Overview**

For all mammals, including humans, vitamins are essential nutrients that are needed in only minute amounts to provide optimum health (Lehninger 1982). We show herein that ascorbic acid (AA) is needed in large amounts and therefore, is not a vitamin, although it is commonly called "vitamin C".  We list many other reasons why this is so.   The "vitamin" identity was mistakenly assigned to AA about 20 years before its 1927 discovery by Nobelist Albert Szent-Gyorgyi who later, with another Nobelist Linus Pauling, both showed AA was not a vitamin (Pauling 1987; Stone 1972).  The mistaken view that AA is a vitamin is immediately apparent from the following: about 4000 mammals synthesize AA, on average circa 150 mg/kg body weight daily (vastly more in stress) or about 10 g/day, normalized to 70 kg "human" weight (see Pauling p.98, line 6 and p.100, line 1). Humans are in a very small subgroup of 5 mammals that has lost the ability to synthesize AA (Chattergee 1973).  However, the human has the same needs per kg as the 4000 normal mammals who still synthesize AA.  If less than that amount of AA is acquired, a rise occurs in incidence of virtually all diseases, including infections and cancer, rapid aging and early death. This is mistakenly considered  normal morbidity and mortality (M&M) by those not aware of the numerous major needs for AA in multigram quantities. This for example occurs in the replacement of structural proteins in resistance to aging and disease.  To explain how such mistaken views could persist so long, we note that in all of past history, and perhaps throughout the future, people will suffer ailments with causes and possibly cures that have no known logical basis or scientific explanation.  Much of medicine is compelled to either accept or deny these empirical associations.  Thus, empiricism has long been a guide of medical practice.

In early mammalian evolution, synthesis of AA evolved to provide this versatile endogenous molecule to control, organize, drive and take part in many vital processes (Stone 1972).  It was only by accident, that a very few mammals in the tropics obtained enough AA in their food to survive a low stress environment without endogenous synthesis.  Freedom from the need to synthesize AA gave a genetic advantage to these animals until they left the tropics.  Pauling (1987, 1991) and others have reported that, rather than milligrams, several grams (over ten in many people) a day of ascorbic acid (AA) are absolutely necessary for numerous aspects of optimum health in humans**.** The confusion over AA's non-vitamin nature has been prolonged by continued use of the misnomer "Vitamin C" for communication convenience by those who knew better (even Pauling).

The large quantities of AA shown above to be synthesized by normal mammals have been found to be essential for optimum health (Chattergee et al 1975; Pauling 87, 91; Ely 1999). When in stress (e.g., pain, fear, disease, etc.), the needs and synthesis are far higher (~100 fold!) (Lewin 1976, p.109).  Such mammals, often called "normal", were here on earth long before humans. It is important to note that, in the wild state, the normal mammals do not get scurvy and do not consume a refined diet (i.e., their free sugar intake is very low).  They developed and used AA as an important endogenous molecule that is necessary in such large quantities for essentially all systems and functions. AA enables all normal mammals to resist most diseases, intoxications, and other disorders of humans including tuberculosis, polio, etc. In the human body, there are thought to be about 50,000 different kinds of enzymes, a large number of which are influenced by AA (Pauling 1987, p.87). As AA levels fall, many enzyme systems fail, disorganizing numerous physiologic functions (King 1967). The human need for AA is the same as that of the normal mammals (Pauling 1987, Chap 9). When this need is met and their blood sugar is maintained at a low level as described later, humans are also protected.

The normal mammals meet the fluctuations in AA need automatically over a wide range of conditions due to the adaptability and efficiency of a global endocrine synchrony. Unfortunately, the human as a non-synthesizing mammal, does not have this synchrony but is burdened with the same daily AA requirements. Thus, humans have the life-long need (*not* an option!) to supplement inefficiently an amount that is unavoidably greater than the basic normal mammal need. Simply providing humans a calculated amount of ~150 mg/kg body weight as one or a small number of doses (oral or other) on a fixed schedule will not replicate the sufficiency seen in normal mammals.

We cite many reports (from a much larger number) demonstrating that high AA is required for optimum or "full health" as defined to be "the greatest resistance to disease" by Albert Szent-Gyorgyi, the Nobel Prize winner who first isolated AA (Stone 1972, p.xi). He also said AA was not a vitamin, and "The medical profession itself took a very narrow and wrong view: Lack of ascorbic acid caused scurvy, so if there was no scurvy there was no lack of ascorbic acid." (Stone 1972, p.xi). Linus Pauling, the only holder of two unshared Nobel Prizes explained at great length why AA is not a vitamin (Pauling 1987, Chap 9). He also estimated that most people get only 2% of the AA necessary for full health (Stone 1972, p.x).

In essence, the benefits of high AA in the synthesizing mammals are so great *and* the countless costs of low AA (due to the "vitamin view") in humans are so tragic that neither could be an accident. To *think* of AA as a vitamin is a lethal error. It dooms the thinker to rapid aging, much illness with related suffering, medical costs, and early death. When AA first became available in the 1930's, its impressive effects in infectious, as well as, degenerative and other diseases stimulated extensive world-wide research and clinical activity resulting in publication of over 10,000 papers by 1975. Hundreds of these were carefully reviewed in the book by Irwin Stone (1972) the biochemist who introduced Pauling to AA. However, successes of AA in clinical trials against colds and cancer in the 1960's and 70's were compromised by the high blood glucose levels accepted as normal in developed nations (Ely 1981, 1996a,b; Santisteban et al 1985; Hamel et al 1986; Ely et al 1988). Until the 1900's, the *low-sugar*, whole-grain, unrefined diet produced 2-hour postprandial blood glucose values of 50-90 mg/deciliter. These are still seen where the primitive diet prevails (Chatterjee and Bannerjee 1979, Tab.1 footnotes) and are approximately half the glycemic levels typical of affluence today (Ely 1996b). Listed below is a sampling (from a much larger number) of ways in which it was shown decades ago that AA at high doses gives "full health". Although it was not understood at that time, it is now known how and why this was so. As described in more detail above, hyperglycemia opposes AA because glucose competitively inhibits the insulin mediated active transport of AA against the high gradient; in health, AA in cells is ~50 times the plasma level (Lewin 1976, p.78). The inability to raise INTRAcellular AA to optimum levels, because of hyperglycemia (GAA), compromises all biochemical processes. The correlations between dietary factors (e.g., high sucrose, meat and animal fat), and incidence and M&M for both heart disease and cancers of various types are so strong, that WHO and numerous agencies and investigators are accepting them as indicating true causality (Carroll 1977, Skegg 2003; Scobie 2002; Ely and Krone, 2002a). New Zealand and the US have some of the highest sugar intakes and cancer rates in the world (Carroll 1977).

By the 1950's, the convenience of numerous antibiotics and vaccines had captured the attention of mainstream medicine and no one felt compelled to read the older literature or practice its methods. AA was soon rejected as unpatentable and forgotten by mainstream medicine. Nevertheless, in the last 50 years, many reports have been published by research oriented physicians of viral and bacterial diseases, intoxications, etc., successfully (safely and economically) treated by massive doses (from 10 to over 200 grams (!) per day) of ascorbate. If a normal mammal is responding to intoxication similar to chemotherapy in a human, the mammal is observed to increase its synthesis of AA so much that it excretes in urine over 100 times as much AA as it normally has in its entire body (Lewin 1976, p 109; Longenecker 1939). High levels of AA are needed in the liver to hydroxylate toxins so they will be excreted in the urine and demethylate the toxins so they are not stored in the body fat (Tsao 1997).

**SECTION 3. Prevention and Cure of Major Disorders of Affluence**

The effects discussed above are manifest and important in almost every bodily function and medical procedure in humans and other mammals. This includes those mammals that are "normal" (i.e., still synthesize their own AA circa 10 g/d normalized to 70 kg body mass) (Pauling 1987, p.98). A sampling is listed below in alphabetical order: aging, antitoxin, birth defects, cancer, cardiovascular disease, cold trial failures, infection and surgery.

**Aging.** The changes associated with aging can be strongly opposed by AA through several mechanisms (Ely and Krone 2002b; Krone and Ely 2004). There is a large daily need of AA for the hydroxylation reactions necessary for synthesis of new connective tissue to maintain its youthful characteristics (Pauling 1987, p. 92). Maintenance of youthful flexibility and elasticity are necessary: (i) for the heart and the arteries to provide maximum ejection fraction with minimum effort and backpressure; (ii) for pulmonary efficiency; and (iii) for leukocyte deformations in diapedesis and phagocytosis; etc. The free radical theory of aging was proposed by Denham Harman in the 1950's (Harman 1969). The antioxidant properties of AA can counteract the free radical damage. Moreover, AA can oppose protein glycation, another commonly cited biochemical change associated with aging (Cerami et al 1987; Ely et al 1988; Krone and Ely, 2004).

**Antitoxin.** Today, we are susceptible to many conditions, frequently fatal, for which high AA has been suggested by both human and animal studies to offer an effective treatment. For example, it was reported to save the patient quickly in many forms of shock (Stone 1972, Chaps 24 and 27). Many thousands of needless deaths occur each year from the bites of snakes, spiders, other insects, etc. Klenner obtained rapid reversal of the swelling, pain, breathing difficulties, shock, etc, resulting from such "bites" using intravenous AA (350 to 700 mg/kg body weight) and AA p.o. as followup (Klenner 1971). Similarly, Klenner normalized the shock in patients with barbiturate poisoning using intravenous ascorbate (12 to 75 g) (Klenner 1971). More than 50 years ago high levels of AA had been shown to prevent or cure intoxications due to organic chemicals (e.g., benzene) and heavy metals (Hg, Pb, As, mercurial diuretics, etc.) (Stone 1972; Chapman 1947). A second modern example of the mechanisms now understood for beneficial effects of AA is that, in Hg intoxication: AA has been found to enter brain cells and act as an electron donor, converting $Hg^{++}$ to $Hg°$, thus mobilizing it and facilitating excretion via sweat (Ely et al 1999a).

**Birth Defects.** Gestational diabetes causes the more frequent of two types of very serious fetal injury arising from hyperglycemic antagonism of mitosis. One type produces major malformations in early and the other neurological defects in late pregnancy. The same mechanism of injury and method of prevention apply in each case. The mechanism, *GAA* (Ely 1981) and its proof are discussed above, and the theory has been cited in a review (Cousins 1983). In essence: (1) normal mitosis requires ribose from the HMS which runs at a rate proportional to intracellular AA, and fetal cells typically have the highest AA concentrations (fetal organs ca 180 mg %) recorded in human tissues (Lewin 1976, p. 78); but (2) hyperglycemia competitively inhibits insulin-mediated active transport of AA into these cells, slowing mitosis. It was also shown that hyperglycemia of early pregnancy induces striking reproductive anomalies in an animal model (Hamel et al 1986). Thus, high levels of AA and low blood sugar can eliminate most birth defects.

**Cancer.** Neoplastic initiation occurs daily even in healthy humans but adequate AA produces basement membrane surrounding each cell that prevents metastasis. Also the wbc will attack the cancer if glucose is low and AA is high leading to high leukocyte intracellular AA. When cancer is being treated with chemotherapeutic agents, AA is needed to prevent the side effects by ensuring that the liver can metabolize these agents, as described above (Tsao 1997).

**Cardiovascular Disease.** To prevent cvd, not only is there a high need for AA in renewal of structural proteins (see above), but also high intakes of AA and E prevent the development of new lesions of the vascular intimal surface, if glycemic control is adequate.

**Cold Trial Failures and rCHO.** The cold trials of the 1960's and 1970's failed because of the high blood sugars prevalent in affluent nations. Blood sugar has a strong tendency to be higher in persons with a high rCHO intake. The trials could be repeated using blood sugar (such as glycated hemoglobin (GHb)) as an independent variable. Then it would be found that persons with very low GHb had low incidence and low severity of colds because their AA was largely intracellular as explained above. Those persons with high GHb (and therefore low intracellular AA) would have high incidence and severity of colds.

**Infectious Diseases 1.** In the 1930's, the remarkable Claus W. Jungeblut, MD, worked in the College of Physicians and Surgeons of Columbia University. He first reported that AA in concentrations, attainable in humans by a high intake, could inactivate and or protect against numerous viral and bacterial pathogens and their toxins (Pauling 1987; p.166; Jungeblut 1935a,b, 1937). These include the polio, hepatitis and herpes viruses, etc. Many other researchers have published in vitro, in vivo, animal and human evidence of AA's almost universal ability (bacteriostatic, bactericidal, virucidal, etc.) to prevent and cure infections. One of the earliest research findings was AA's ability to neutralize and render harmless many bacterial toxins (e.g., tetanus, diphtheria, and staph toxins) (Stone 1972; Jungeblut 1935b, 1937). It has been clearly demonstrated that oral AA 2 g daily protects against post-transfusion hepatitis (Pauling 1987). The high safety, efficacy and economy of AA have been proven, and low levels of AA greatly predispose to infectious diseases (Jungeblut 1935a,b, 1937; McCormick 1951).

**Infectious Diseases 2.** Another ingenious physician-researcher, Fred R. Klenner, MD (from Duke Medical School 1936), was originally a chemist and a PhD student in Physiology. He was inspired by Jungeblut and the fact that AA is an essential nutrient, *not* a vitamin. As a result, he perfected high dose AA therapy and used it for many infectious diseases and other disorders (Klenner 1949, 1971, 1974). He reported that early dosing with amounts that elevate wbc AA sufficiently (50-100 g/d by iv and/or oral)** produces prompt (3-7 day) cures of polio, viral encephalitis, acute hepatitis (all types), adenovirus, chicken pox, measles etc. Robert F. Cathcart, MD, world famous for his work in orthopedic surgery at Stanford, adopted and extended the work of Klenner, treating over 30,000 patients. His innovations include a simple method called "titrating to bowel tolerance" (TBT) dosing that permits a very sick outpatient to administer AA in exactly the correct oral dose each day for influenza, glandular fever, etc (Cathcart 1981, 1984). Treatment protocols and results for high-dose AA have been reported in detail by Klenner (1949, 1971, 1974), Hoffer (1971, 1991), and Cathcart (1981, 1984).
** In iv dosing, AA is always sodium ascorbate.

The amount of AA that humans can get from food alone is more than adequate for prevention of TB but not pertussis and far below that required to treat these diseases. As per the literature, TB is one of the easiest diseases to prevent because the AA bacteriostatic level is only 1 mg% (i.e., even below the AA renal threshold 1.4 mg%) (Boissevain 1937). The AA bactericidal level for TB was shown to be 10 mg% by Sirsi (1952). But, TB patients require much higher AA doses to maintain a given AA level than is observed in other diseases. It is surprising that nearly all of the many authors reporting (in 1930's and 1940,s) on use of "vitamin like" AA doses stated it improved TB patients, but they never tried multi-gram doses (cost may have been a factor). Of course, as explained earlier, if their AA-synthesizing ability had not been lost, humans would be immune to TB without AA dosing.

**Surgery.** For years, surgeons have expressed much interest in factors that could facilitate wound strength, healing rapidity, and freedom from infection. All three of these conditions are strongly influenced by ascorbic acid (AA). Recently, a large scale ongoing clinical study of critically ill surgical patients has shown many benefits in those given multi-gram AA and vitamin E (Nathens et al 2002). Yet, it has been known for 60 years (Bourne et al, 1946; Bartlett 1942) that scar tissue is much stronger in guinea pigs and humans with high intake of AA; the strength of the scar tissue increasing linearly with the AA dose. In 1941, it was shown that human unhealing wounds of 7 months duration healed rapidly when 1 g AA/day was given for 10 days (Lund 1941). And, medicine as a whole has shown little interest in the fact that patient AA is commonly low at surgery and falls precipitously in surgery, trauma or other stress (Pauling 1987; Stone 1972; Crandon et al 1958). Hume reported the dangerous fall of white blood cell AA to scorbutic levels within 12 hours after heart attack in 31 consecutive patients (Hume et al 1972). When AA is used at high levels, vitamin E need is increased because it prevents clotting without causing a hemorrhagic tendency (Ochsner et al 1950; Ochsner 1964).

No valid argument or experimental evidence supports the concept that "AA is a vitamin". The fact that AA is not a vitamin has been proven countless times.

**SUMMARY AND CONCLUSIONS**

As cited above, it is established by unquestioned authorities including CDC and WHO that over 1 million patients receiving TOC have suffered and died annually for decades in the US. Included in this annual M&M were approximately 700,000 cvd, 500,000 cancer, 200,000 infection and 150,000 stroke patients. Although physicists and other scientists who perished in this toll were capable of understanding the erroneous beliefs of organized medicine explained herein, they may have trustingly assumed TOC's were the best modalities. Physicists et al who have read this far, likely now understand the errors of TOC's and the rationale of the highly effective therapies that had been rejected as unprofitable. Likely, most or all of these scientists now recognize the real peril that they, their families and all of us are suffering today. It is expected most of these physicists will urge the UCS to cite this LANL in the UCS Newsletter to stimulate action to disempower the Medical Quality Assurance and Medical Disciplinary Boards and the TOC's they enforce.

See Chapter 2, "Dental Mercury Amalgams and the Biochemical Relationship with Micromercurialism, Retention Toxicity and Extreme Psychoses (Serial Killers with No-Remorse)", below these references.

**REFERENCES**

Angell M. Is academic medicine for sale? New Engl J Med 2000;342(20):1516. www.nejm.com/content/2000/0342/0020/1516.asp)

Augustin LS, Polesel J, Bosetti C, Kendall CW, La Vecchia C, Parpinel M, et al. Dietary glycemic index, glycemic load and ovarian cancer risk: a case-control study in Italy. Ann Oncol 2003 Jan;14(1):78-84.

Bartlett MK, Jones FM, Ryan AE. Vitamin C and wound healing: ascorbic acid content and tensile strength of healing wounds in human being. New Engl J Med 1942;226:474-481.

Boissevain CH, Spillane JH. Effect of synthetic ascorbic acid on the growth of tuberculosis bacillus. Am Rev Tuberculosis 1937;35:661-62.

Bourne GH. The effect of vitamin C on the healing of wounds. Proc Nutr Soc 1946;4:204-11.

Cathcart RF. Vitamin C, titrating to bowel tolerance, anascorbemia, and acute induced scurvy. Med Hypotheses 1981;7:1359-76.

Cathcart RF. Vitamin C in the treatment of acquired immune deficiency syndrome (AIDS). Med Hypotheses 1984;14(4):423-33.

Carroll KK. Dietary factors in hormone-dependent cancers. Curr Concepts Nutr 1977; 6:25-40

Centers for Disease Control National Vital Statistics System. Deaths, percent of total deaths, and death rates for the 15 leading causes of death in 10-year age groups, by race and sex: United States,1999. Table LCWK2. September 2001. www.cdc.gov/nchs/data/lcwk2.pdf

Cerami A, Vlassare H, Brownlee M: Glucose and aging. Scientific American, 1987; 256(5): 90-96.

Chapman W, Shaffer CF. Mercurial diuretics. Arch lntern Med 1947;79:449-456.

Chatterjee IB, Bannerjee A. Estimation of dehydroascorbic acid in blood of diabetic patients. Anal Biochem 1979;98:368–74.
Chatterjee IB: Evolution and the biosynthesis of ascorbic acid. Science 182: 1271-1272, 1973.
Chatterjee IB, Majumder AK, Nandi BK, Subramanian N. Synthesis and some major functions of vitamin C in animals. Ann NY Acad Sci 1975;258:24-47.
Cheraskin E, Ringsdorf WM Jr: How much refined carbohydrate should we eat? Am Lab July 1974; 6:31-35.
Cooper MR, McCall CE, DeChatelet LR. Stimulation of leukocyte hexose monophosphate shunt activity by ascorbic acid. Infect Immun 1971;3:851-853.
Cousins L. Congenital anomalies among infants of diabetic mothers: etiology, prevention, prenatal diagnosis. Am J Obstet Gynecol 1983; 147: 333–338.
Crandon JH, Landau B, Mikal S, et al. Ascorbic acid economy in surgical patients as indicated by blood ascorbic acid levels. New Engl J Med 1958;258:105-113.
Ely JTA. Hyperglycemia and major congenital anomalies. New Engl J Med 1981;305:833.
Ely JTA. Glycemic modulation of tumor tolerance. J Orthomolecular Med 1996a;11(1):23-34.
Ely JTA. Unrecognized pandemic subclinical diabetes of the affluent nations: causes, cost and prevention. J Orthomolecular Med 1996b;11:95-99.
Ely JTA. Ascorbic Acid and Some Other Modern Analogs of the Germ Theory. J Orthomol Med 1999;14:143-156.
Ely JTA. Inadequate levels of essential nutrients in developed nations as a risk factor for disease: a review. Rev Environ Health. 2003 Apr-Jun;18(2):111-29.
Ely JTA, Krone CA. Glucose and Cancer. New Zealand Med J, 2002a;115(1159):U123
Ely JTA, Krone CA. Aging: predictions of a new perspective on old data. Exp Biol Med 2002b Dec; 227(12): 939-42.
Ely JTA. Warner GA, Read DH, Santisteban GA. Protein glycation: ascorbate antagonism [abstract] Bull Am Physical Soc 1988; 33 : 296.
Ely JTA, Fudenberg HH, Muirhead RJ, LaMarche MG, Krone CA, Buscher D, Stern EA. Urine mercury in micromercurialism: bimodal distribution and diagnostic implications. Bull Environ Contam Toxicol 1999;63:553-9.
Feltman J, editor. Prevention's Giant Book of Health Facts. Emmaus PA: Rodale Press, 1991, p. 352.
Grimble RF.Effect of antioxidative vitamins on immune function with clinical applications. Int J Vitam Nutr Res 1997;67(5):312-20.
Hamel EE, Santisteban GA, Ely JTA, Read DH. Hyperglycemia and reproductive defects in non-diabetic gravidas: a mouse model test of a new theory. Life Sci 1986;39:1425-28.
Harman D. Prolongation of life: role of free radical reactions in aging. J Am Geriatr Soc 1969;17(8):721-35.
Harman D. The aging process. Proc Natl Acad Sci USA 1981;78(11):7124-28.
Harman D. The biologic clock: the mitochondria? J Am Geriatr Soc 1972;20(4):145-7.
Hems G. The contributions of diet and childbearing to breast cancer rates. Br J Cancer 1978;37:974-82.
Hoffer A. Ascorbic acid and toxicity. New Engl J Med 1971;285:635-36.
Hoffer A. Clinical procedures in treating terminally ill cancer patients with vitamin C. J Orthomolecular Med 1991;6:155-160.
Hume R, Weyers W, Rowan T, Reid DS, Hillis WS. Leucocyte ascorbic acid levels after acute myocardial infarction. Brit Heart J. 1972; 34: 238-243.
Hutchinson ML, Lee WYL, Chen MS, Davis KA, Ely JTA, Labbe RF. Effects of glucose

and select pharmacologic agents on leukocyte ascorbic acid levels, Fed. Proc. 1983; 42:930 (abst).

Jungeblut CW. Inactivation of poliomyelitis virus by crystalline vitamin C (ascorbic acid). J Exper Med 1935a;62:317-21.

Jungeblut CW, Zwemer RL. Inactivation of diphtheria toxin in vivo and in vitro by crystalline vitamin C (ascorbic acid). Proc Soc Exper Biol Med 1935b;32:1229-34.

Jungeblut CW. Inactivation of tetanus toxin by crystalline vitamin C (ascorbic acid). J Immunol 1937;33:203-214.

King CG. Ascorbic acid (Vitamin C) and scurvy. In: Williams RJ , Lansford EM, editors. Encyclopedia of Biochemistry. New York: Reinhold Pub; 1967, p. 95-99.

Klenner FR. The treatment of poliomyelitis and other virus diseases with vitamin C. Southern Med. Surg 1949;111:209-214.

Klenner FR. Observations on the dose and administration of ascorbic acid when employed beyond the range of a vitamin in human pathology. J Appl Nutr 1971;23:61-68.

Klenner FR. Significance of high daily intake of ascorbic acid in preventive medicine. J Prev Med 1974 Spring:45-69.

Koroljow S. Insulin coma therapy. Psychiatric Quarterly 1962; 36: 261-70.

Krone CA, Ely JTA. Vitamin C and glycohemoglobin revisited. Clin Chem 2001;47(1):148.

Krone CA, Ely JT. Ascorbic acid, glycation, glycohemoglobin and aging. Med Hypotheses. 2004 Feb;62(2):275-9.

Lederberg J. Infectious History. Science 2000 14 Apr;288:287 - 293.

Lehninger AL. Principles of Biochemistry. New York: Worth Publ. 1982, p. 250.

Lewin S. Vitamin C: Its molecular biology and medical potential. New York: Academic Press; 1976.

Longenecker HE, Musulin RR, Tully RH, King CG. An acceleration of vitamin C synthesis and excretion by feeding organic compounds to rats. J Biol Chem 1939; 129:445-453.

Lund CC, Crandon JH. Human experimental scurvy and the relation of vitamin C deficiency to postoperative pneumonia and to wound healing. JAMA 1941; 116(8);663-68.

McCormick WJ. Vitamin C in the prophylaxis and therapy of infectious diseases. Arch Pediat 1951;68:1 -9.

Michaud DS, Liu S, Giovannucci E, Willett WC, Colditz GA, Fuchs CS. Dietary sugar, glycemic load, and pancreatic cancer risk in a prospective study. J Natl Cancer Inst 2002 Sep 4;94(17):1293-300.

Murata A, Morishige F, Yamaguchi H. Prolongation of survival times of terminal cancer patients by administration of large doses of ascorbate. International Journal for Vitamin and Nutrition Research Suppl 1982;23:103-113.

Nathens AB, Neff MJ, Jurkovich GJ, Klotz P, Farver K, Ruzinski JT, Radella F, Garcia I, Maier RV. Randomized, prospective trial of antioxidant supplementation in critically ill surgical patients. Ann Surg 2002 Dec;236(6):814-22

Nilsen TI, Vatten LJ.Prospective study of colorectal cancer risk and physical activity, diabetes, blood glucose and BMI: exploring the hyperinsulinaemia hypothesis. Br J Cancer 2001 Feb 2;84(3):417-22.

Ochsner A. Thrombembolism. New Engl J Med 1964; 271: 211.

Ochsner A, Debakey ME, Decamp PT. Venous thrombosis. JAMA 1950; 144:831-34.

Pauling LJ. Orthomolecular Psychiatry. *Science* 1969;160:265-71.

Pauling LJ. How to live longer and feel better *New York: Avon Books*, 1987.
Pauling LJ. Case Report: Lysine/ascorbate-related amelioration of angina pectoris. J Orthomolecular Med 1991;6(3-4):144-46.
Penn ND, Purkins L, Kelleher J, Heatley RV, Mascie-Taylor BH, Belfield PW. The effect of dietary supplementation with vitamins A, C and E on cell-mediated immune function in elderly long-stay patients: a randomized controlled trial. Age Ageing 1991 May;20(3):169-74.
Price WR. The present status of our knowledge of the relation of mouth infection to systemic disease. Dent Rev 1917;31(4):271-298.
Relman AS. Reforming our health care system: a physician's perspective. Phi Beta Kappa Key Reporter 1992;58(3):3-5.
Riordan N H, Riordan HD, Meng X, Li Y, Jackson JA. Intravenous ascorbate as a tumor cytotoxic chemotherapeutic agent. Medical Hypotheses 1995;44:207-213.
Rivers JM. Safety of high-level vitamin C ingestion. Int J Vitam Nutr Res Suppl. 1989;30:95-102.
Santisteban GA, Ely JTA, Hamel EE, Read DH, Kozawa SM. Glycemic modulation of tumor tolerance in a mouse model of breast cancer. Biochem Biophys Res Commun 1985;132(3):1174-1179.
Schoen RE, Tangen CM, Kuller LH, Burke GL, Cushman M, Tracy RP, Dobs A, Savage PJ. Increased blood glucose and insulin, body size, and incident colorectal cancer. J Natl Cancer Inst 1999 Jul 7;91(13):1147-54.
Scobie B. Cancer flourishes. NZ Med J 2002, 115:304.
Sirsi M. Antimicrobial action of vitamin C on M. tuberculosis and some other pathogenic organisms. Indian J Med Sci (Bombay) 1952;6:252-55.
Skegg DCG, McCredie MRE. Comparison of cancer mortality and incidence in New Zealand and Australia. NZ Med J 2003; 115:205-208.
Stevens CW, Glatstein E. Beware the Medical-Industrial Complex. The Oncologist 1996;1(4):IV-V.
Stone I. The Healing Factor. Vitamin C Against Disease. New York: Grosset & Dunlap, 1972, 152-171.
Trowbridge JP, Walker M. The yeast syndrome. New York: Bantam, 1986, pp. 72-77.
Tsao CS. An overview of ascorbic acid chemistry and biochemistry. In: Packer L, Fuchs J, eds. Vitamin C in health and disease. New York: Marcel Dekker, Inc 1997:25-58
Watson TF Jr. Health service. J Am Soc Prev Dent 1973;3:12-13.
WHO. World Health Report 2000. ISBN 924156198X June 2000. (http://www.who.int/whr/2000/en/index.html)

**CHAPTER 2: Physical Characteristics and Effects Reported for Some High-Copper Dental Mercury Amalgams and the Biochemical Relationship with Micromercurialism, Retention Toxicity and Extreme Psychoses (Serial Killers with No-Remorse)**

**INTRODUCTION**

The so-called "Amalgam Wars" have gone on for close to 200 years and controversy regarding safety of dental amalgam continues to circulate among government regulatory/legislative bodies, the dental/medical community and the general public. This is despite the undisputed fact that Hg is a potent neurotoxin and the mounting research evidence in humans that dental amalgam fillings release significant amounts of Hg vapor and persons with larger numbers of fillings have greater body burdens of this toxic element. Herein, we mention briefly physical/chemical factors that can affect mercury release from dental amalgam (DA) (e.g., chemical composition, intermetallic forms, mechanical stresses, temperature, acidity, etc). Also discussed is a proper model of mercury toxicity created and accepted by world experts. The model makes clear how and why any person might enter a state called micromercurialism (MM) without any idiosyncratic predisposition. Dental amalgam Hg has been cited as the cause of intoxications that range from the almost imperceptible (tremor, memory, fertility) through severe illness in dental patients (Alzheimer disease and motor neuron disease), to suicidal behavior and serial killers with no-remorse (Bundy, Green River, OKC-Bomber, etc, etc). Dag Brune and other experts have performed research studies that demonstrate conditions that suggest millions of people likely are suffering each day because of these neglected amalgam toxicities.

**THE AMALGAM WARS.**

Mercury dental amalgams in restorative dentistry have faced opposition for nearly 200 years. The first use of mercury amalgam as a dental filling occurred in the 1820's in Paris France. These first formulations tended to expand upon solidification and led to fractured teeth. The Crawcour brothers from Great Britain began a dental practice in New York City in 1833 despite the fact that they were not dentists. They advocated a "new" amalgam of silver and Hg as a cheaper and easier-to-apply substitute for gold fillings. The Crawcours were very successful and competing dentists assembled a vigorous opposition to mercury fillings, in large part because the fillings caused very serious mercurial poisonings, diseases of the gums, etc. Several American associations for dentists banned the use of dental amalgam and even suspended members for malpractice for using mercury amalgam. However, some continued to experiment with "silver" fillings and worked to achieve the proper balance between the silver (which leads to expansion of the amalgam) and tin (which leads to contraction). By the beginning of the 1900's, a "stable" formulation had been developed and waning opposition seemed to have ended this "First Amalgam War"

The "Second Amalgam War" began in ~1926 when the German chemist, Professor A. Stock, began a warning campaign against mercury. Micromercurialism (MM) is a term coined by Stock in the 1920's and widely used to denote chronic intoxication from long term exposure to low levels of Hg vapor. Stock also demonstrated (in himself) that MM can be caused by dental amalgams.  Stock stimulated a wave of scientific interest in MM in Germany and Russia (which adopted the lowest tolerable work place Hg vapor levels in the world, 10 and 1 $\mu g/m^3$, respectively) (Patterson et al. 1985; Gerstner and Huff 1977).  Stock's campaign was  overshadowed by events of World War II in Europe and it was not until the early 1980's that Swedish scientists including Jaro Pleva, Magnus Nylander, Mats Hanson and many others initiated the "Third Amalgam War".  This war continues today with professional dental societies in the US maintaining their position that Hg amalgams are "safe and effective" in spite of the now overwhelming evidence that Hg is released from dental amalgam and accumulates in the tissues of persons with Hg dental fillings (Guzzi et al. 2006).

**THE CHEMICAL/PHYSICAL NATURE OF AMALGAM AND HG VAPOR RELEASE**

The composition of the most common dental amalgams are intended to have about 50% Hg.  Silver and tin are the other major components.  The addition of copper (up to 30%) in recent times has increased in prevalence.  In general, DA is formed by mixing Hg with the other elements (Pleva 1989, 1994). Although a wide range of Cu content (from about 10 to over 30%) has been used in high-Cu DA for the last two decades or so, several of the more popular have about 12% Cu (instead of the formerly more common 3%). Two reasons suggested for the development of these "high-Cu" amalgams were the better structural characteristics (smoother finish and longer holding of edges) and the lower cost of Cu versus Ag.  It is reported that the ADA owns the patents for the high-Cu formulation used by most dentists in the US (Null and Feldman 2002, p.101).

However, an extremely hazardous unforeseen outcome resulted from the high-Cu formulations.  Rather than creating a more stable DA, the high-Cu formulations under the conditions in the human mouth emit much more Hg vapor than conventional amalgam (Brune et al. 1983).  A typical property of the modern high-Cu DA is the "sweating out" of Hg droplets (Pleva 1994).  It has been reported that one of the widely used 12% Cu amalgams releases Hg about 50 times as fast as the 3% Cu (Brune et al. 1983); see Abstract, Fig. 4, and page 70, top paragraph in right column.  It is trivial to demonstrate with a Jerome 411 Hg Vapor Analyzer that high Cu amalgam does release Hg much faster (Ely 2001b).  In hundreds of subjects, it was found that even when > 10 years old these amalgams commonly released Hg $\geq 2$ times the rate of the conventional 3% Cu.

**MERCURY TOXICITY AND THE PROPER MODEL**

We describe a proper model of mercury created and accepted by world experts with no financial interests in dental amalgam; these include Dag Brune, Boyd Haley, Joachim Mutter; Magnus Nylander, Victor Penzer, Jaro Pleva, Patrick Stortebecker, etc.  The model makes clear how and why any person might enter a state called micromercurialism (MM) without any idiosyncratic predisposition.  Initially, essentially every human has some chronic exposure to ambient environmental Hg that does not exceed the person's excretory ability.  If the chronic exposure rises due to any cause (workplace, dental amalgams, etc), a level of intake can be reached that exceeds the subject's excretory ability.  Then, according to the model, the accumulation of Hg is not simply the difference between intake and the original excretory ability, but rapidly becomes much worse, approaching the total intake. The theoretical reason is that the rising soft tissue Hg burden inactivates the enzymes involved in excretory processes (presumably by bonding as Hg++ to SH groups). It is well known that Hg strongly inhibits enzymes (Webb 1963), which can lead to greatly diminished excretion, a condition now called "retention toxicity", and the associated diagnostic difficulties (i.e., low urine Hg excretion; <5 µg/day). Diagnosis of MM frequently hinges mistakenly on urine Hg values intended to detect acute intoxication (i.e., normal values (0-20 µg/L) and "toxicity" (>150 µg/L)). A 1938 Public Health Service study of MM in hatters published in JAMA, showed that those with the highest body burdens and who were the most intoxicated exhibited the lowest urine Hg (Neal and Jones 1938), and thus were in "retention toxicity". According to the proper model, most MM subjects initially exhibit retention toxicity and very low urine Hg (i.e., 0-5 µg/d).

People who excrete Hg well do not retain it and can tolerate large intakes, maintaining low blood levels and preventing diffusion of Hg into bone storage. In retention toxicity, the decreased excretion and rise in blood levels are associated with diffusion of Hg into the skeleton where large body burdens can accumulate in bone storage (Bloch and Shapiro 1981). It has also long been known from x-ray fluorescence studies that persons with chronic Hg exposure such as dentists, have elevated skeletal Hg stores (Bloch and Shapiro 1981), as well as, low urine Hg (Neal and Jones 1938). If the source of exposure is eliminated (environmental or amalgams), subjects recover from retention toxicity and exhibit elevated Hg excretion, sometimes for many years from bone storage (in agreement with both the proper model and measurements by world experts). Moreover, in old age, osteoporosis may accelerate Hg release to a rate somewhat above that corresponding to the 20 year excretion half-time estimated by Sugita (1978). As a result of these complexities, there is almost no awareness of the degree of Hg intoxication in the US today or its impact on the incidence of Alzheimer's disease (AD), Parkinsons, suicide, psychoses, etc.

The diagnosis of MM is always missed by most physicians who misinterpret the resulting low (or "unmeasurable") urine Hg (Ely et al 1999) as indicating no intoxication. Low urine Hg is MULTIPLY MISTAKENLY interpreted today almost universally as evidence of low intoxication attributed to low exposure, also falsely attributed to careful Hg-handling practices in the clinics. In a US study of 6925 dentists, 90% had urine Hg concentration below 6 µg/L (Echeverria 2002). This study and it's very low value have been cited many times and misinterpreted to indicate very low intoxication and low exposure due to (allegedly) high levels of control in dental clinic atmosphere**. In fact, anyone with amalgams or occupational exposure will have urine levels two or more times above this low value unless in retention toxicity (urine Hg <5µg/day). The few physicians who diagnose Hg intoxication correctly realize such impairment of excretion will worsen until exposure stops or the patient dies.

**Editors of a dental "journal" have been quoted as claiming that "amalgam is a stable alloy similar to sodium chloride" (found by Mutter et al. 2004a, p.393).

The "Expert" Problem. Intoxication is a normal concern of physicians who understand and can treat acute intoxication by Hg as occurs in lab/industrial spills, etc. In recent decades medicine has decreased use of Hg in medicinals but has relied on the dental profession to police the amalgam Hg toxicity question. In 1991, the FDA and the AMA sided with the American Dental Association on the supposed safety of amalgam and all called for more research to allay public fears on the issue. However, Stortebecker had shown and warned that Hg levels in the mouth (and in brain by "shortest pathway") due to dental amalgam were much higher than occupationally permitted (Committee on Mercury Compounds 1969). MM differs in many respects from the generally recognized acute form of Hg intoxication. These differences, which were recently elucidated (Ely et al. 1999), are responsible for almost universal failure of today's physicians to diagnose MM. Thus, it is unfair to expect anyone with today's training in medicine to be aware of the basics on MM.

**DISORDERS, INCLUDING PSYCHOSES, ASSOCIATED WITH MICORMERCURIALISM**

The MM disorders are believed (by research scholars) to range from the innocuous (such as hand tremor) to serious neurological diseases (including ALS, Alzheimer's, MS and Parkinson's) (Stortebecker 1986) and to the frank psychoses of the hatters, light-house keepers, and Newton whose madness ended one of the most productive scientific careers in history (Klawans 1991). Several studies have reported links between suicidal behavior and occupational mercury exposure (e.g., dentists and mercury miners), as well as, in subjects exposed only to amalgam Hg (Guzzi et al. 2006; Ametz et al 1987; Grum et al 2004). Dental amalgam (DA) Hg may even be associated with serial killers characterized by "no-remorse" (i.e., Green River Killer, OKC bomber, Post Office Massacres, Unibomber, etc). Much evidence that supports the seemingly bizarre (but easily testable) hypothesis that the wave of unprovoked homicidal behavior (characterized by no-remorse) (that some say is sweeping the US) may have, dental Hg as a major factor. We propose that this has become much more probable with the increasing popularity of the "high copper" type of dental amalgams.

Obviously, not all individuals who receive one or more of the 100 million amalgams that are placed each year in the US become no-remorse killers. However, there are a number of factors that can greatly influence the magnitude of the body burden and symptoms (including no-remorse) that result from amalgam Hg. It has long been evident from the literature that prolonged (chronic) exposure of humans and animals to extremely low levels of mercury (Hg) and or lead (Pb) result in storage of these metals in the skeleton** from which they slowly leak out to the brain causing a great variety of neurological and psychiatric disorders. If only Hg is involved, chronic intoxication is called MM. At the present time, widespread use of the newer "high-copper" amalgams (that release mercury 50 times as rapidly) has changed the intoxication to an extreme form of MM.

**(at many times the brain levels that incapacitate in acute exposure).

Although Hg alone at levels attainable in the body from amalgams can cause psychoses or other intoxication, very much smaller amounts of Hg when combined with even smaller amounts of Pb have been reported to produce toxic effects including death in animal models. This Pb-Hg Synergism was discovered by Schubert (1978) who reported that a very small dose of Hg (LD1, the dose that should kill only 1% of the animals) plus an even smaller dose of Pb (1/20 LD1) exert an amazing synergism (possibly via enzyme inactivation) to produce an LD100 effect killing all the animals! Evidence of the same synergism was recently reported in humans by Godfrey and Campbell (1994). In spite of low Pb gasoline, militia and others who regularly use firearms still have Pb exposure from firing ranges as well as Hg from amalgams and may be at risk for this synergism.

Many people do not absorb sufficient Hg systemically from their dental amalgams to cause overt disease. However, there are three factors that differ so greatly from person to person that, in theory, two individuals with the same amalgam geometry can differ in the Hg absorbed by a factor greater than 100,000 fold! Estimates suggest that only 5 to 10% of the amalgam wearing population may be at significant risk for serious disease. In addition to the greater release of Hg from the high copper amalgams, two physiological factors can greatly enhance Hg exposure. Mouth breathing (i.e., sleeping, exercising, etc., with the mouth open) can increase absorption 4000 times! Clenching, bruxing, or chewing increases release rate about ten fold. Thus, according to published testable figures, absorption can be increased 400,000 times in a person with all three of these factors (Ely 2001a).

**THE FAILURE OF MEDICINE AND DENTISTRY**

The "Mercury Wars" have occurred because the level of science in the US was unable to prepare medicine and dentistry to understand MM and two of its principal characteristics (retention toxicity and bone storage). It is unfortunate (although *certainly* understandable) that physicians refused to supervise the medical aspects of amalgams. Naturally, medicine wasn't interested in tooth fillings. Physicians aren't aware of these matters. For example, physicians in general and even most dentists don't understand that lower urine Hg indicates higher mercury poisoning. So, again the physicians have not been prepared to be a help to this cause.

**DENTOLOGY**

It is proposed that there be a specialized branch of medicine (dentology). The students will have the same curriculum of courses that normally constitute medical school training. It is conceivable that students can transfer during pre-med years and possibly even during medical school with approval of the institution. In a similar fashion, the interns may transfer with approval of the departments concerned.

**TESTS OF THE HYPOTHESIS.**

Uneducated public feeling against those who commit loathsome acts of violence is understandably punitive and would be against tests that might show that the "criminals' behavior" was or might have been due to intoxication arising in the manner considered here. However, if the hypothesis is correct the very safety of the country is at risk; we cannot afford to allow this possibility to go unstudied. To test the Hg no-remorse hypothesis would involve determining: (1) if the subject currently has, or has had in the past, amalgam dental fillings; (2) the type of amalgam (e.g., high copper); (3) the number of amalgam fillings; and (4) presence of the other susceptibility factors mentioned above. If it appears from this initial screening that Hg might be implicated, further medical testing would be warranted. The placing 100 million amalgams per year in the US should be investigated and stopped.

**REFERENCES**

Arnetz BB, Horte LG, Hadberg A, et al. (2003) Suicide among Sweedish dentists: a ten-year follow-up study. Scand J Soc Med. 15:243-46.

Bloch P, Shapiro IM (1981) An x-ray fluorescence technique to measure the mercury burden of dentists in vivo. Med Phys 8:308-311

Brune D, Gjerdet N, Paulsen G (1983) Gastrointestinal and in vitro release of copper, cadmium, indium, mercury and zinc from conventional and copper-rich amalgams. Scand J Dent Res 91:66-71

Committee on Mercury Compounds. (1969) Maximum allowable concentrations of mercury compounds. Arch Environ Health 19(6):891-905.

Echeverria D (2002) Mercury and dentists. Occup Environ Med 59(5):285-6

Ely JTA, Fudenberg HH, Muirhead RJ, LaMarche MG, Krone CA, Buscher D, Stern EA (1999) Urine mercury in micromercurialism: bimodal distribution and diagnostic implications. Bull Environ Contam Toxicol 63(5):553-9

Ely JTA (2001a) Risk factors for parenteral intoxication by mercury from dental amalgam. Bull Environ Contam Toxicol 67(3):309-16

Ely JTA (2001b) Mercury induced Alzheimer's disease: accelerating incidence? Bull Environ Contam Toxicol. 67(6):800-6

Ely JTA. Letter to US Atty Gen Janet Reno. Nov 10, 1996. (unpublished)

Ferracane J, Adey J, Wiltbank K, et al. (1999) Vaporization of Hg from Hg-in amalgams during setting and after abrasion. Dent Mater. 15(3):191-5.

Gerstner HB, Huff JE (1977) Clinical toxicology of mercury. J Toxicol and Environ Health 2:491-526

Godfrey M, Campbell N. (1994) Confirmation of mercury retention and toxicity using DMPS. J Adv Med 7:19-30.

Grum DK, Arneric N, Kobal AB et al. (2004) Can occupational expoure to elementary mercury increase the risk of suicide? RMZ Materials Geoenviron. 51:452-57.

Guzzi G, Grandi M, Cattaneo C, Calza S, et al. (2006) Dental amalgam and mercury levels in autopsy tissue. Food for thought. Am J Forensic Med Pathol. 27:42-45.

Huggins HA. It's all in your head. Life Sci Press:Seattle, 1989, pp235.

Klawans H. Newton's Madness. N.Y. Harpers 1991.

Koike M, Ferrance JL, Adey JD, et al. (2004) Initial mercury evaporation from experimental Ag-Sn-Cu amalgams containing Pd. Biomaterials Jul;25(16):3147-53

Lewin S. (1976) Vitamin C. Its Molecular Biology and Medical Potential. New York Academic Press. 1976; 231.

Mutter J, Naumann J, Sadaghiani C, Walach H, Drasch G. (2004a) Amalgam studies: disregarding basic principles of mercury toxicity. Int J Hyg Environ Health. Sep;207(4):391-7.

Mutter J, Naumann J, Sadaghiani C, Schneider R, Walach H. (2004b) Alzheimer disease: mercury as pathogenetic factor and apolipoprotein E as a moderator. Neuro Endocrinol Lett. 25(5):331-9.

Neal PA, Jones R (1938) Chronic mercurialism in the hatter's fur-cutting industry. JAMA 110:337-3433

Null G, Feldman M. (2002) Mercury dental amalgams: The controversy continues. J Orthomolecular Med 17(2):85-110.

Nylander M (1986) Mercury in pituitary glands of dentists. Lancet i:442

Patterson JE, Weissberg BG, Dennison PJ (1985) Mercury in human breath from dental amalgams Bull Environ Contam Toxicol 34:459-68

Pendergrass JC, Haley BE (1997) Inhibition of brain tubulin-GTP interactions by mercury: Similarity to observations in Alzheimer's diseased brains. In: Sigel A, Sigel H (eds) Metal ions in biological systems, Vol 34. M. Dekker, New York, pp 461-478

Penzer V. Research needed re amalgam safety. ADA News, July 30, 1984. in: Huggins HA. It's all in your head. Life Sci Press:Seattle, WA, 1989; p.234.

Pleva J (1994) Dental mercury-A public health hazard. Rev Environ Health 10:1-27

Pleva J (1989) Corrosion and mercury release from dental amalgam. J Orthomolec Med 4(3):141-158.

Schubert J, Riley EJ, Tyler SA. (1978) Combined effects in toxicology - A rapid systematic testing procedure: cadmium, mercury, and lead. J Tox Env Health 4: 763-776

Stortebecker P (1986) Mercury poisoning from dental amalgam - A hazard to human brain. Bio-Probe, Inc, Orlando, FL

Sugita M. The biological half-time of heavy metals (1978) The existence of a third, "slowest" component. Int Arch Occup Environ Health. 41(1):25-40.

Thompson CM, Markesbery WR, Ehmann WD, Mao XY, Vance DE (1988) Regional brain trace-element studies in Alzheimer's disease. Neurotoxicol 9:1-7

Webb JL (1963) Enzyme and metabolic inhibitors. Academic Press, NY